# Fundamental Dilemmas in Theoretical Physics

## Hisham Ghassib
ghassib@psut.edu.jo

## The Princess Sumaya University for Technology

## Amman, Jordan.
## 2014

**Abstract**

In this paper, we argue that there are foundational dilemmas in theoretical physics related to the concept of reality and the nature of mathematics in physics. Physical theory is treated as a conceptual organism which develops under the weight of its internal contradictions.

The paper discusses in depth the problem of objective reality in physics and its relation to scientific practice. Then, it explores the problematic relation between physical meaning and mathematics in modern physical theory, followed by a discussion of the trend of contemporary physics to replace physical principles with pure mathematical principles. Finally, it discusses the problem of logical coherence in modern physical theory. The paper emphasizes the importance of resolving these dilemmas to the proper practice of theoretical physics.

**Introduction**

There is a growing feeling among an increasing number of theoretical physicists that theoretical physics is in deep structural crisis. Basically, it is not a specific concrete crisis related to a set of concrete theoretical and practical problems, but a deep structural crisis lying at the very heart of theoretical physics as an enterprise. It could threaten the whole enterprise of theoretical physics.

In particular, in quantum mechanics, the very concept of objective reality is questioned and the dominant interpretation tends to negate the concept of a pure objective reality, and to accord the subjective observer a privileged role in determining reality on a random basis. It will be argued, on methodological bases, that this negation endangers the whole enterprise of theoretical physics.

Secondly, it will be argued that the primacy and priority of abstract mathematics in 20th century theoretical physics problematize the very possibility of physical meaning in physics. In this regard, modern physics, epitomized by special relativity, general relativity, black hole physics and the various formulations of quantum mechanics, is thoroughly contrasted with classical physics, particularly as regards concept formation in both.

Thirdly, the problem of theoretical complexity and logical coherence in physical theory is addressed and highlighted, in view of the problem of infinities and singularities in quantum field theory and modern cosmological theories.

One of the tools used to investigate these major dilemmas is the view, previously worked out by the author (Ghassib, 1988; 2012a; 2012b), that physical theory is an organism, rather than a set of loosely connected partial theories. This view turns out to be essential for appreciating the importance and reality of the dilemmas referred to above.

The author believes that some of the fundamental concrete problems in contemporary theoretical physics would be intractable without a thorough investigation of the dilemmas referred to.

This paper will consist of five broad sections. Section 1 will deal with the meaning of sickness in theoretical physics and its contemporary symptoms. Section 2 will deal with the problem of ideology and dogma in theoretical physics, in relation to the problem of reality. Section 3 will deal with the problematic relationship between mathematics and physical meaning. Section 4 will deal with the problematic relationship between physical principles and mathematical principles. The last section will deal with the problem of logical coherence in physical theory.

## 1.    Physics as Organism:

Modern Physics does not seem to be immune to dogma, ideology and even bigotry (Ghassib, 1993; 2003; Maxwell, 2008). These defects have played a noticeable role in conceptualizing theoretical physics, particularly in 20[th] century physics (Jammer, 1966), even though the two mechanisms of critique and experimentation tend to limit this role and shorten its duration of efficacy. This in itself is a form of structural and fundamental tension and crisis. Physics boasts of its objectivity and its growing appropriation of reality. The three defects mentioned clearly express subjectivity of the worst kind. Thus, they can be considered a form of malady or sickness afflicting the body of theoretical physics (Ghassib, 2012a; 2012b).

This characterization implies the idea that physics is a living organism, which grows under the pressure of its internal and external contradictions. It is indeed so, as I have shown in a previous work ( Ghassib, 1988).

Theoretical physics seems to suffer fundamentally from a number of structural maladies which express real dilemmas that urgently need to be addressed. These dilemmas do not primarily consist of the inability of physical theory to appropriate and explain new experiments, as was the case at the turn of the 20[th] century (Heisenberg, 1949; Kragh, 1999). Rather, they are fundamental and internal conceptual and methodological dilemmas that affect the very enterprise of theoretical physics. There are at least four major domains in modern theoretical physics, each of which is a site of a major fundamental dilemma.

## 2.    Ideology and the Problem of Reality:

Physics has erected its glory and credibility as reliable knowledge, and has fought its battles with dogma and despotism, on the basis of the notion of an objective reality, which can be known using varied creative scientific rationality that is the climax of rational thought (Ghassib, 2012b; 2012c). Physics has had to follow a very tortuous and laborious path to establish the notion of objective reality in modern culture.

Yet, we find a tendency in modern physics to transcend the notion of objective reality, and even negate and deny it, without pondering the enormous methodological, epistemological and ethical ramifications of such a move. Such anti-realists behave as though the many cultural battles fought by physics in the last few centuries were mere empty bubbles devoid of all meaning and value.

Quite a few of the top quantum physicists have gone out of their way to deconstruct the notion of objective reality and negate its necessity in physics and elsewhere (Grib, 2013; Dorato, 2014; Pusey, 2012; Briggs, 2013; Squires, 1996; Mermin, 1985; Rogers, 1999).

Niels Bohr seems to adopt a neo-Kantian stance, whereby physics cannot, in virtue of its basic methodology, have access to an objective reality, and can only have access to phenomena within the context of humanly constructed experimental set-ups, thereby opening the domain of physics to magic, mysticism, and religion. ( Cale, 2002; Tanona, 2002; Degen, 1989; Cuffaro, 2011).

Heisenberg promotes the notion of potential reality, which is a virtual space of probabilities, which can be transformed into actuality by observation and measurement, thus, endowing this human-all-too-human operation with the power to create being out of nothing (Heisenberg, 1949; 1971; 1979).

Eugene Wigner, Von Neumann and Pauli went so far as to consider human consciousness a necessary condition of the collapse of the wave function, and, thus, a condition of concrete material being (Wigner, 1961; 1967; Jammer, 1966).

Major physicists and mathematicians, such as Von Neumann and Dirac, embarked on inventing an elaborate mathematical scheme to ground this basically ideological dogma and support it with mathematical proofs and axioms. (Birkhoff, 1936; von Neumann,1955).

When Louis de Broglie, the founder of wave mechanics, challenged these subjectivist idealist interpretations of quantum mechanics with his objective interpretation, known as the double solution or pilot wave interpretation (De Broglie, 1923; 1924; 1927; 1987), at the 1927 Fifth Solvay Conference (Valentini, 2010), he was subjected to a sustained barrage of ridicule and verbal abuse by the Copenhagen "gang", which succeeded in silencing him for twenty-five years at least. Only after David Bohn rediscovered his interpretation in 1952 (Bohm, 1952) did he dare to resume his work on his objective interpretation.

This is not to say that the De Broglie-Bohn interpretation is entirely satisfactory or even as satisfactory as the Copenhagen family of interpretations, but the point is that the suppression of De Broglie's interpretation in the way it was ruthlessly suppressed cannot be explained and justified on a purely scientific basis. It was basically an ideological philosophical choice, and in fact in recent decades, quite a few studies have been conducted, which have shown the extent to which that choice was ideological and philosophical (Lee, 2006; Cuffaro,2011).

What should be pondered here is, why was De Broglie not given the opportunity to defend his interpretive scheme and to defend a conception of reality that physics had struggled for centuries to establish and sustain? How are we to explain the passionate vehemence with which De Broglie was confronted in 1927, and with which Einstein himself was confronted on more than one occasion following 1927 (First, 2012; Norsen, 2005) ? It is, of course, ideology, and philosophical dogma-- the usual struggle between subjectivism and realism, idealism and materialism.

Ever since the Scientific Revolution in the 17[th] Century, physics has been a thorn in the flesh of ideology and philosophical dogma. As a result, the latter has tended to modify its tenets to accommodate itself to this scientific challenge. However, with the advent of quantum mechanics and the Copenhagen interpretation of quantum mechanics, an attempt has been made to reverse the issue, whereby subjectivist ideology would attempt to appropriate fundamental physics and use it as weapon against competing ideologies. What is at stake in all these battles is the concept of objective, mind-independent, reality.

In this regard, the Einstein-Bohr debate was the basis of subsequent debates (Jammer, 1974; Bohr, 1949; Shimony, 1993).

Einstein courageously and brilliantly challenged the Copenhagen clique on that particular issue, as he realized what it means to abandon this concept or negate it or neutralize it. He refused to surrender to their, particularly Bohr's, alluring reasoning and to their ideological tactics (Whitaker, 2006; Ballentine, 1972; Stachel, 1983; Einstein, 1948). His tenacity was interpreted, wrongly, as a reactionary response and prejudice bordering on senility (Fine, 1986).

Instead of listening carefully to what Einstein was saying, and to cooperate with him to find a satisfactory way to rehabilitate the concept of objective reality, they were concerned only with combating, refuting, rebuffing and ridiculing his arguments. It was obviously an ideological war.

The question that was deliberately absented from the Copenhagen discourse was: What prompted Einstein so stubbornly to stick to the concept of objective reality and defend it? Why did Einstein deem this concept a necessary condition for any meaningful physical theory, and perhaps, for other concerns? These questions could be more broadly posed as follows: Why did Einstein and why do we insist on the integrity of the concept of objective reality? Why do we stubbornly stick to it? Why do we resist abandoning it, even when we are prepared to abandon classical physics as such? Why are we prepared to abandon certain essential features of classical physics, but not prepared to abandon this ontological basis?.

The full answer could be quite elaborate and multi-disciplinary, and would of necessity go beyond the scope of this paper. Thus, we will give here a partial answer as follows.

Our basic thesis here is that scientific practice as such presupposes objective reality (Ghassib, 2012b).

Thus, the interpretations, which negate or condone it, enter into contradiction with the very practice which they purport to be produced by it. Thus, with these interpretations, scientific practice entangles itself into a self-contradiction, which could lead to the demise of the whole scientific enterprise. It is a form of epistemological suicide.

It is our contention that scientific practice presupposes scientific rationality. The basis of the latter is: to confirm the existence of a significant phenomenon, to specify it via physical quantities, which are in turn fixed via mathematics and measurement, and to demonstrate their mechanisms of emergence and the process of this emergence (Bhaskar, 1978).

The essence of this rationality is the dialectical relationship between mathematized theorization and precise experimentation. (Ghassib, 2012c).

In this sense, scientific rationality presupposes that nature is a physical system consisting of an infinite collection of physical subsystems which interact with each other. These interactions are the only causal sources of physical phenomena. These subsystems are unified via a network of physical quantities, which are related to each other in various ways. These diverse relations govern the way these physical quantities vary. Scientific practice has no sense and no concrete purpose without this ontological picture of nature and natural phenomena.

Thus, the Copenhagen interpretation seems to blow up the very basis, meaning, purpose and raison d'etre of scientific practice.

It is also to be noticed that the Copenhagen interpretation--or, interpretations, to be more precise-- emanates from the act of measurement. This choice is rooted in positivism and its noticeable impact on theoretical physics in the first third of the 20$^{th}$ century.

Positivism regards the act of measurement the pivot of scientific practice and the central determining component of physical experience.

However, this choice is not an inevitable choice. One could as well start from experimentally validated theory by considering the latter the essence of knowledge, the axis of scientific practice and the determining factor of physical experience. That was precisely what was done by David Bohm, at least in his 1952 paper ( Bohm, 1952).

Bohm did not start from the act of measuring macroscopic quantities, as Bohr and Heisenberg specifically had done, but started from the Schrodinger equation, from which he derived two equations, which were amenable to a causal objectivist interpretation. This suggests that philosophy plays an important role in the interpretations of quantum mechanics. Thus, whereas the Copenhagen interpretations presuppose an amalgam of positivism and subjective idealism, the 1952 Bohm interpretation presupposes a form of critical realism (Bhaskar, 1978). The problem is that the Copenhagen interpretations have tended to conceal their philosophical presuppositions and give the impression that they follow necessarily from purely physical experiments and ideas.

## **3.** **The Problem of Meaning in Physics:**

The essential point here is that, in classical physics, physical meaning comes before mathematical representation and precedes it logically. Accordingly, mathematical representation comes as a culmination of the process of meaning generation and synthesis. For example, the Newtonian concept of force has a determinate meaning prior to its mathematical representation and prior to Newton's equations of motion. Its meaning follows from Newton's conception of the Universe and the principle of causality. It is an expression of material interaction between material particles in absolute space and absolute time. It is the basic cause of change. Its mathematical properties and its role in the laws of nature and in physical equations follow specifically form this definite physical meaning. Similarly, the classical concepts of electric and magnetic fields have a definite physical meaning prior logically and historically prior to the formulation of Maxwell's equations. The latter came as an expression and extension of their meaning-- as a crowning moment and completion of this meaning. (Ghassib, 2012b).

Besides, in classical physics, the laws of nature logically precede the natural phenomena to be explained, and constitute a basis for explanations.

Let us compare these features of classical physics with corresponding features of quantum mechanics. Let us start with wave mechanics. It is clear here that the discovery of the wave function and the specification of its physical meaning did not precede the discovery of the Schrodinger equation. The latter came first and the attempts to specify the meaning of the wave function have come later in a persisting on-going process. (Kragh, 1999).

The mathematical representation in this case precedes the physical meaning, and it even gives the impression that it replaces it. It is some sort of a neo-Pythagoreanism which tends to deem the mathematical equation the only objective reality. The rest are partially or totally subjective.

Also, the relationship between the laws of nature and natural phenomena seems to be turned upside down in quantum physics. Here, the laws of nature are no longer a basis for explaining phenomena, but, rather, phenomena acquire the function of tools for exploring the physical meaning of physical equations.

The same observations apply to Heisenberg's Matrix Mechanics and Wigner's distribution function (Wigner, 1932; Ghassib, 1996; 2012b).

The equation comes first, and the never-ending quest for physical meaning commences afterwards. The problem is that this neo-Pythagorean transformation opens  physics wide open to irrational subjectivist interpretations. The very logic of scientific discovery has been turned upside down, endangering its very rationality.

## 4. <u>Physical Principles versus Mathematical Principles:</u>

A keen observer of the history of 20[th] century physics would readily notice a trend whereby mathematical principles are rapidly replacing physical principles. This is most conspicuously evident in the development of Einstein's research project.

Einstein started his scientific career with general universal physical principles, which are unchanging and absolute (Ghassib, 2010; Pais, 1982).

He did not derive these universal principles from mathematical considerations, but from a new method for reading familiar experiments. That was precisely what Einstein did in formulating the principle of special relativity, the principle of the constancy of the speed of light, the equivalence principle, and, to a certain extent, the principle of general relativity.

As we know, Einstein was prompted by these principles to revise the very notions of space and time and the very notion of spacetime geometry (Einstein, 1905). However, in the process of transition from special relativity to general relativity, he increasingly fell under the spell of mathematics, and turned from a search for physical principles to a search for mathematical principles ( Einstein, 1949; 1954; Pais, 1982). Accordingly, his field equations came as a result of a dialectical synthesis between physical principles and mathematical principles (Ghassib, 1999).

Following this grand  synthesis, the mathematical tendency grew rapidly, in Einstein and in theoretical physics generally, until it has become overwhelmingly hegemonic, whereby mathematical principles have almost completely replaced universal physical principles.
Thus, we have seen Einstein spending the last thirty years of his life in search of very abstract geometric principles for constructing a comprehensive unified field theory that would explain gravity, electromagnetism and matter, and would replace quantum mechanics and general relativity (Sauer, 2007). Is his failure to arrive at such a theory related to his shift from the physical to the mathematical?

This shift was not confined to Einstein, but proved to be a continuing general trend in theoretical physics. This is noticeable in such theories as string theory, quantum gravity, loop quantum gravity and supergravity, and in such principles as gauge symmetry, duality principles and the holographic principle.

These are highly abstract mathematical principles, with almost no empirical content-- very unlike the older physical principles, which had a conspicuous physical meaning organically related to experiment. Theoretical physicists have been turned into mere applied mathematicians and calculating machines.

## **5.** **The Question of Logical Coherence in Physical Theory:**

If theory is to be a mathematically deductive system capable of providing definite testable results, it must possess a high degree of logical coherence and self-consistency (Ghassib, 2012a).

However, historically speaking, no physical theory is perfectly logically coherent and self-consistent. All physical theories possess a degree of logical inhomogeneity. However, it is noticeable that, as physical theories have become more elaborate, theoretical as opposed to empirical, complex and ambitious, this element of logical incoherence has become more conspicuous and threatening, which prompts the questions: Is there a limit to the complexity and scope of physical theory? Is theorization intrinsically limited? Do physical theories collapse under their own weights? Do they grow in complexity until they reach a critical point of collapse? Would an affirmative answer to these questions entail a limit to systematic, scientific knowledge?

This phenomenon of logical incoherence to the point of collapse has reached its zenith in quantum field theory and quantum gravity. Physics has found ways to get round these incoherences, but it has not overcome them. The problems of divergences in quantum field theory have been mellowed using normalization tricks, but only partially and not entirely satisfactorily. String theories and quantum gravity theories may have solved some of these incoherence problems, but at the expense of losing touch with the empirical world (Smolin, 2006).

Will physics find a way to reverse this trend? Or, is scientific knowledge intrinsically limited? Or, is the universe partially knowable only? These are not mere philosophical questions, but genuinely physical questions, which must be faced by the theoretical physics community. The future of their discipline hangs on finding satisfactory answers to these dilemmas.

## **Conclusion:**

In this paper, we have drawn attention to, and elaborated, four foundational and structural dilemmas in theoretical physics, related to the question of objective reality, the relationship between mathematics and physical meaning, the relationship between physical principles and mathematical principles, and the questions of logical coherence in physical theory. We have highlighted the crucial relevance of these questions for the practice of theoretical physics. This should be understood as a call to the theoretical physics community to pause and ponder the very foundations of their practice.